# Adhesive Release via Elasto-Osmotic Stress Driven Surface Instability

Khulood Al-sakkaf[2], Monicka Kullappan[1,3] and Manoj K. Chaudhury[1,2]

[1]Department of Chemical and Biomolecular Engineering

[2]Department of Materials Science and Engineering, Lehigh University

Bethlehem, Pennsylvania 18015, United States

[3]Current Address: GE Research, Niskayuna, New York 12309. United States

**Abstract.** Recent studies demonstrated that an elastomer containing hygroscopic inclusions absorbs moisture and swell. Here we show that a thin film of such an elastomer bonded to a rigid substrate undergoes morphological instability upon absorption of water, the wavelength of which increases linearly with its thickness. As the driving force for such a morphological instability arises from the difference of the chemical potential of water between its source and that in the film, its development is slowed down as the salinity of the water increases. Nonetheless, the wavelength of the fully developed morphology, but not its amplitude, is independent of the salinity. We also demonstrate that if a domed disk-shaped adherent is attached to the hygro-elastomeric film before moisture absorption, the elastic force generated during the morphological transition is able to dislodge it completely without the need of any external force. These patterns, once developed in pure water, is subdued when the salinity of water increases or if it is exposed to dry air. They re-emerge when the film is immersed in water again. Such an active response could be important in fouling release when a ship coated with such a hygro-elastomer changes its location during its long travel through sea, where salinity varies from place to place.

**Introduction.** Two philosophies have evolved over the years in controlling biofouling of ship's hulls. Firstly, a coating has been sought for many years, the surface of which would totally resist biofouling.[1,2] Secondly, a coating has been sought that would provide easy release of foulants from its surface. While the roles of surface chemistry and mechanics guided scientific and technological developments in this field, and laboratory tests[3-5] have been carried out to validate various models[6-9], identification of a coating that meets the above goals is still far from our sight. An ideal combination[10] of these two approaches would be such that not only fouling would be deterred as much as possible, but also that the adhesion of foulants to the coating would be so weak that the weight of the foulant or the hydrodynamic forces created by the ship's motion would dislodge the biofoulants. The intent of this report is to embark upon a new approach in that the coating itself would undergo a morphological transition in water, so that an *internal repulsive force is generated at the interface of the coating and the foulant,* which would, thus, push the foulant out of the coating. This approach is philosophically similar to an idea described recently[11], where the authors induced an instability thus leading to a crater like pattern on the surface of an electro-active elastomer. Liu et al.[12] showed that a thin film undergoing a wrinkling instability due to an applied electrical voltage is also capable of releasing biofoulants from its surface. These demonstrations of active biofouling control using a dynamic topography has many interesting possibilities. What we describe here is an attempt to achieve a similar result using an osmotically triggered hydrostatic instability.

Fracture mechanics teaches us that the stress applied externally to a bonded system is amplified at the crack tip, which is characterized by a square root singularity. In the real biofouling release situations, however, the above condition is not always met and thus the singularity is tempered[13] (Figure 1). For example, if the coating is much smaller than the dimension of the foulant, which is typically the case with many foulants, the stress intensity factor is reduced by their ratio.[9] This reduction can be so huge that release becomes virtually unrealistic. Secondly, the adhesive released by the fouling organism can spread on the coating forming a thin film ahead of the thicker central region of the foulant.[3-5] Here, crack does not initiate from the edge of contact, so that the only venue to release is the central region, where crack initiates as bubbles via interfacial cavitation. Following cavitation, the bubbles would nucleate and a crack would be formed that then would eventually propagate from the center to the outer region of contact as a run away instability. This mode of failure requires unduly high stress, which, while is beneficial to achieving strong gecko-mimetic adhesion[14], is a detriment for easy release. What is needed in moving forward is to design a coating that would generate a repulsive force all along the interface, thereby pushing the foulant off the surface of the coating.

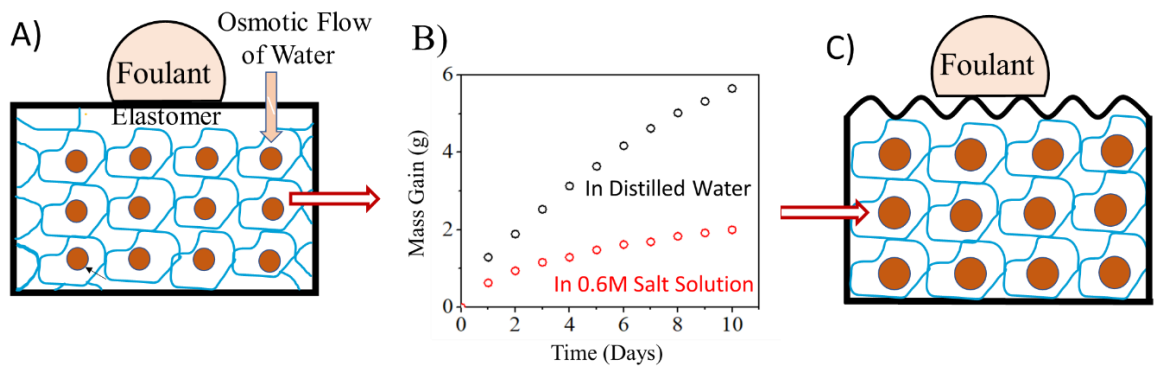

**Figure 1.** Schematic representation of a foulant in contact with a hygroscopic elastomer (A), which absorbs water from the surrounding (A) and undergoes a morphological instability (C) thereby pushing the foulant out of the surface. (C) Mass of water gained by such a hygroscopic elastomer (see text) depends on the molarity of the surrounding aqueous phase. These data were obtained with strips (56 mm x 23 mm x 2 mm) of H-PDMS elastomer immersed in distilled water and 0.6 M aqueous salt solution.

*The main idea* of this research is described in Figure 1. Here, the coating itself is made of an elastomer, in which hydrophilic components are dispersed. When exposed to water, such a coating would swell by absorbing water from the surroundings. As the droplets grow inside the elastomer, two possibilities can emerge. In one scenario, the entire composite film would undergo a volume expansion. Since the film is thin and is well bonded to a rigid support, such a volume expansion would be possible only with a homogeneous increase of its thickness. On the other hand, such a homogeneous deformation is intrinsically unstable with a thin film, as a perturbation would lead to an inhomogeneous growth, thus leading to an Morphological instability in soft elastomeric films. Various types of morphological patterns have been observed in numerous settings, which would be too dauting to review in this paper. Interested readers may consult the reviews cited in references 6-9 and the references cited therein. In a previous study involving thin bonded elastomeric films[15], such type of instability was observed when a flat rigid object was released from the film with the application of an external force. Various analysis [6-9, 15] have shown that the formation of periodic undulation of the film thickness is preferable over a homogeneous deformation as the total elastic energy stored in the undulated film is lesser than that of a homogeneous deformation.

**Hygroscopic PDMS Elastomer.** Modification of elastomers via liquid has recently been investigated by various groups.[16-28] The work reported here is related particularly to those studies concerning hydrophilic liquid inclusions.[16-20] These materials are unique in that they maintain, overall, the intrinsic properties of the elastomer but they are capable of absorbing moisture from the surrounding environment. Initially, glycerol[17--20] was dispersed in PDMS, which has the remarkable advantage that the toughness of the elastomer is increased.[20] More recently, the authors[21] used a hydrophilic polymer, such as polyacrylamide as the hydrophilic inclusion. Idea wise, while our approach to prepare hygro-elastomer is similar to those of the previous studies[21], the base composition of our formulation is a concentrated aqueous solution of fructose and glucose, which is found in abundance in natural nectars, e.g. agave, honey and maple syrup to name a few.[29] Highly concentrated aqueous solution of fructose allows plants to retain moisture in arid environments. Since fructose is present in plants at a high concentration, it is prone to crystallization, which is inhibited by glucose that introduces defects against crystallization. We recently reported[30] that a drop of a concentrated aqueous solution of fructose and glucose, which mimicked the composition of agave, grows several times its original size when exposed to a humid atmosphere, but it shrinks back to its original size when exposed to a dry atmosphere. In this work, we dispersed such a solution in a crosslinkable polydimethylsiloxane (PDMS) liquid along with a hydrophilic polymer (e.g. polyvinyl alcohol or polyacrylic acid) and glycerol. While both the polymers yielded similar results, most of the studies presented here were obtained with polyacrylic acid, which is based on a conjecture that it affords a greater water absorption ability due to its extended H-bonding capacity[24,31] In future, we plan to present the detailed results obtained with these and other hydrophilic polymers. After adding the aqueous solutions of the above components to crosslinkable PDMS liquid, it was stirred vigorously in the presence of a small amount of a surfactant (silicone glycol) as a compatibilizer. After casting a thin film of this mixture on a glass slide coated with a home-made adhesion primer, it was cured at 70 °C. The cured elastomer is considerably softer (elastic modulus 0.28 MPa) than that (2 MPa) of the unmodified PDMS. The resultant film, nonetheless, behaved like a soft rubber film, which exhibited significant resistance to tear while it was repeatedly stretched and released by hand. This observation qualitatively concurs with that reported previously.[20,21] The elastic modulus of the coating, however, increased slightly (0.34 to 0.44 MPa), when it was immersed in water.

**Promoting Adhesion Between Hygroscopic Elastomer to Glass** When the cured film was submerged in distilled water or in an aqueous salt solution, it absorbed water at different amounts and at different rates. The hydrostatic stress in the coating resulting from the swelling of the hydrophilic inclusion eventually led to the morphological instability, which was photographed for subsequent analysis. One problem that arose was that the coating delaminated easily from the substrate (glass slide) due to poor interfacial adhesion. Instability developed in these weakly adhered films are somewhat related to that reported in a beautiful study by Velankar et al[32] , in which folding induced delamination of swelled film from a substrate was observed. In order to carry out the studies reported in this work, strong interfacial adhesion was required. After finding that the commercially available adhesion primers fail to meet the strict demands of the experimental conditions employed here, especially when the glass coated film was immersed in water, we needed to develop our own adhesion primer.

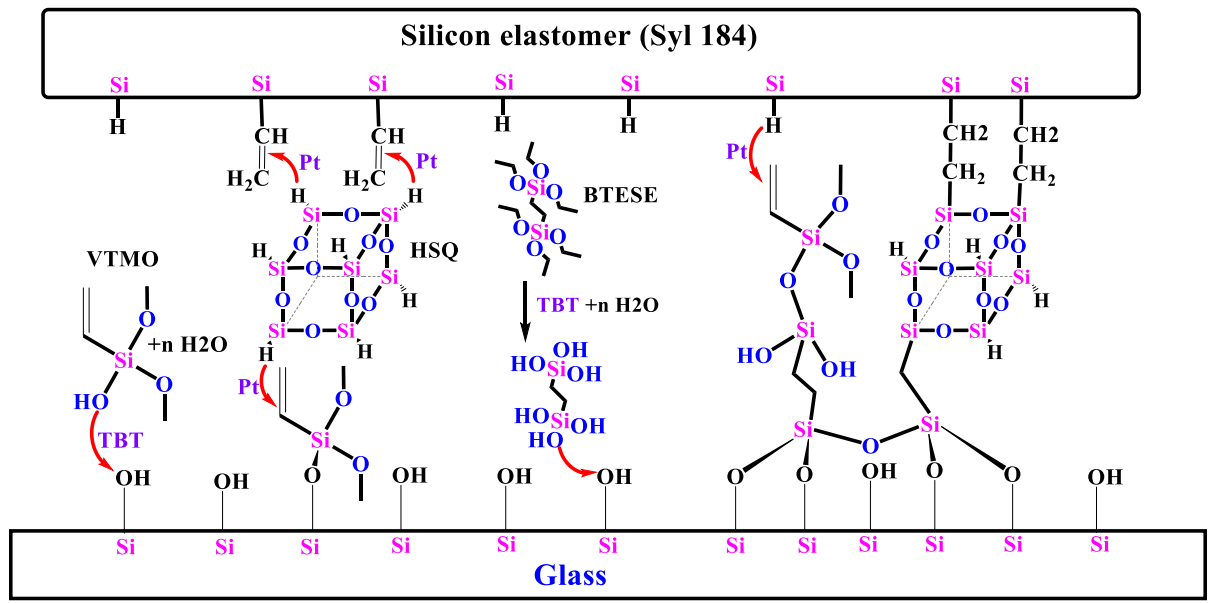

**Figure 2.** Schematic illustration of the synthesis of an adhesion primer for enhancing the strength of adhesion of a hydrosilylated elastomer (e.g. Syl 184) to a glass substrate. The primer is composed of vinyltrimethoxysilane (VTMO), Hydrogen silsesquioxane (HSQ), 1,2 Bis(triethoxysilyl) ethane (BTESE), Titanium-n-butoxide (TBT) and platinum (Pt) catalyst.

This homemade primer uses the basic principles that we reported in an old patent[33], but its composition is different in many ways. In particular, the primer used in this work consisted of a mixture of hydrogen silsesquioxane, vinyl trimethoxy silane, hydrogen silsesquioxane, 1,2 Bis(triethoxysilyl) ethane, titanium-n-butoxide and a platinum catalyst. This multifunctional primer bonds to glass through the alkoxysilane groups via hydrolysis and condensation, which are present abundantly in the primer. The silicone hydride (-SiH) and vinyl groups of the primer can bond to the vinyl and the silicone hydride (-SiH) groups of the PDMS matrix in the presence of the platinum catalyst, as the primer continues to crosslink and grow its network structure. Titanium-n-butoxide ensures that the alkoxysilanes of the primer composition can hydrolyze by reacting with moisture, and promote the reaction between the silanol groups among themselves as well as with other silanol and other alkoxysilane groups.[32,33,34] The primer made from the mixtures of various components is dissolved in chloroform, which has a shelf life of at least a month. However, in the experiment to be reported below, the primer was spin coated on clean glass slides soon after it was prepared (see experimental section). The schematic of Figure 2 depicts our view of how such a primer could improve adhesion of the coating to the glass substrate.

The adhesion between the crosslinked PDMS film and glass achieved in this manner is so strong that cohesive failure occurred when attempt was made to delaminate it from the glass by pushing a blade though the adhesive junction. The adhesive joint was also stable when immersed in distilled and salt water at room temperature, and even when the water was boiled or sonicated. All the studies reported below were performed at room temperature ($22^{o}C$).

The thermodynamic driving force for water diffusing into the hygro-elastomeric PDMS (H-PDMS) stems from the difference of the chemical potential of water between that of the surrounding phase and the inclusion. As expected, this force is highest when the surrounding phase is distilled water, but it decreases with the increase of the salinity of the aqueous phase. Thus, the film exhibits various degrees of swelling depending upon the salinity of the aqueous phase, which varied from zero to that of seawater in our experiments.

**Characterization of Morphological Instability.** Morphological instability ensues when the hygro-elastomeric coating absorbs sufficient water and undergoes volume dilation, which occurs when the coating is exposed to humid atmosphere or when it is kept immersed in liquid water. Figure 3 shows the typical patterns that are formed on the surface of such coatings of various thicknesses after they are immersed in pure water or a salt solution.

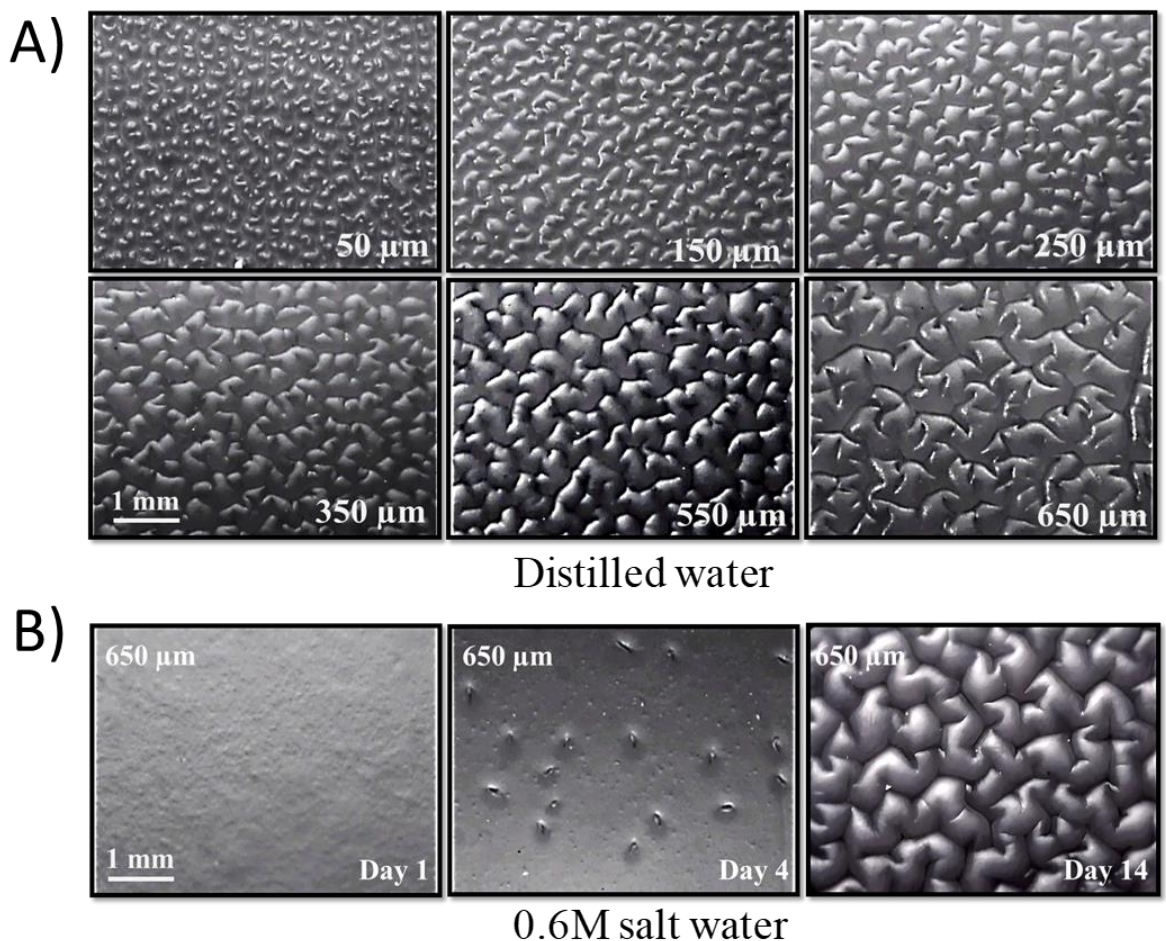

**Figure 3.** (A) Micrographs of the morphological instabilities that developed on H-PDMS films of various thicknesses after they were submerged in distilled water for 24h and 0.6 M salt water for up to 14 days . (B) The thickness of the film for the salt water test is 650 $\mu m$, where the fully formed pattern is preceded by the formation of random spots on the surface.

Patterns develop in saline water more slowly than in pure water. An example of which is presented in Figure 3, where a 650 $\mu m$ thick H-PDMS film was submerged in 0.6 M salt solution. In this case, evolution of small random spots on the surface preceded the development of complete pattern.

There are two length scales[8,9,14,33], *wavelength* and *amplitude* that characterize the observed morphologies. The wavelength of the instability was estimated from the fast Fourier Transform of the images, which were obtained using light microscopy. The data summarized in Figure 4 show that the wavelength ($\lambda$) is rather independent of the salinity of the aqueous phase as well as the hydrophilic content of the H-PDMS, or if prepared either with polyacrylic acid or polyvinylalcohol as a sub-component. $\lambda$ depends only on the thickness ($h$) of the coating, that too linearly, i.e. $\lambda \approx$ 2.6 $h$. The prefactor (2.6) of the thickness is close to, but slightly lower than what was observed (*3.8*) previously[9,35] with PDMS elastomeric films, when a rigid flat object adhering to those were removed by external forces. The exclusive dependence of the wavelength on the thickness, however, indicates that it results from the minimization of the elastic energy due to the longitudinal and the transverse shear deformations of the film as a periodic perturbation develops on its surface. Note that this occurs here with a positive hydrostatic stress in the film that contrasts the previous observations where the hydrostatic stress was negative in the punch pull-out mode.

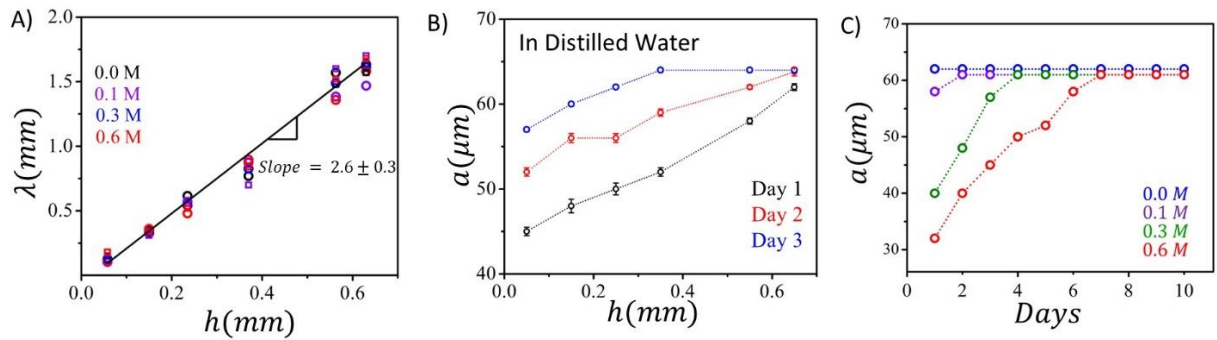

**Figure 4.** (A) Wavelength (λ) of the morphological instability that develops on H-PDMS film as a function of its thickness (*h*) when it is immersed in distilled and with salt water of different salinities. The symbols (i) ○, ○, ○ and ○ ; (ii) △, △, △, and △ correspond to the films synthesized with (i) PAA and (ii) PVA as sub-components in the hydrophilic inclusions (see text). These studies were conducted with the films immersed in 0.0M (distilled water) and 0.1M, 0.3M and 0.6M salt concentrations. The distilled water and 0.1 M data were obtained after 24hrs of immersion, whereas the data for 0.3M and 0.6M were obtained after seven days of immersion (B) Amplitude of the instability in H-PDMS films increases with its thickness. These set of data correspond to immersion in distilled water. (C) Evolution of amplitude for a 600 mm thick film at different salt concentrations.

A *remarkable* feature of the morphological instability is that it is reversible, i.e., when pronounced instability develops in pure water, it disappears when exposed to dry atmosphere, or is diminished in salt water. Then again, when the flattened film is re-immersed in distilled water, the morphological patterns re-emerge fully, Figure 7(G-J). We conjecture that this phenomenon would be relevant for the release of foulants from ship hull, as the movement of the ship through marine water of variable salinities is expected to lead to dynamic transition of the coating's morphology thus inducing a time dependent deformation of the film.

In the context of release of a foulant from the coating, both the normal as well as shear displacement of the surface of the film relative to that of the foulant should facilitate release. The amplitude of instability, nonetheless, would be the main parameter that is directly relevant to the generation of elastic repulsive force needed to dislodge a hard foulant from the coating. These amplitudes were estimated using a Gwyddion V 2.60 software[36-38] (see experimental section), which is designed to analyze the height fields of structures captured via micrographic methods. The software allows implementation of filtering feature based on the wavelet transform and inverted wavelet transform of the two-dimensional set of data acquired using microscopy. The results summarized in Figures 4 (B and C) show that the amplitude of instability increases with thickness as well as with the immersion time. Furthermore, the amplitude increases more rapidly in distilled water than in salt solutions systematically. In all cases, the amplitude appears to saturate to a value close to 60 $\mu$m after a long immersion time. Next, we present some preliminary results of a study in which a domed disk-shaped object strongly adhered to a H-PDMS coating was released in air and under water.

**Adhesion Studies using a Model Adherent.** A domed disk-shaped epoxy, with leaflet like protrusion from its base (Figure 5) was used as a model adherent. Such a domed disk makes flat contact with the coating and mimics the geometry of many natural foulants, which have bulky central structures, but they adhere to a substrate by releasing adhesives[5] that spread as thin films ahead of the central bulky region. With such a geometry, crack has to first nucleate from the central region of contact and then propagate outward against another pinning force arising from the bending deformation of the leaflet structure. All these pose a challenging condition for release in response to external perturbations. In order to provide proof of the concept as depicted in Figure 1, we carried out various types of detachment studies using a 650 µm thick as PDMS film with and without the hydrophilic inclusions.

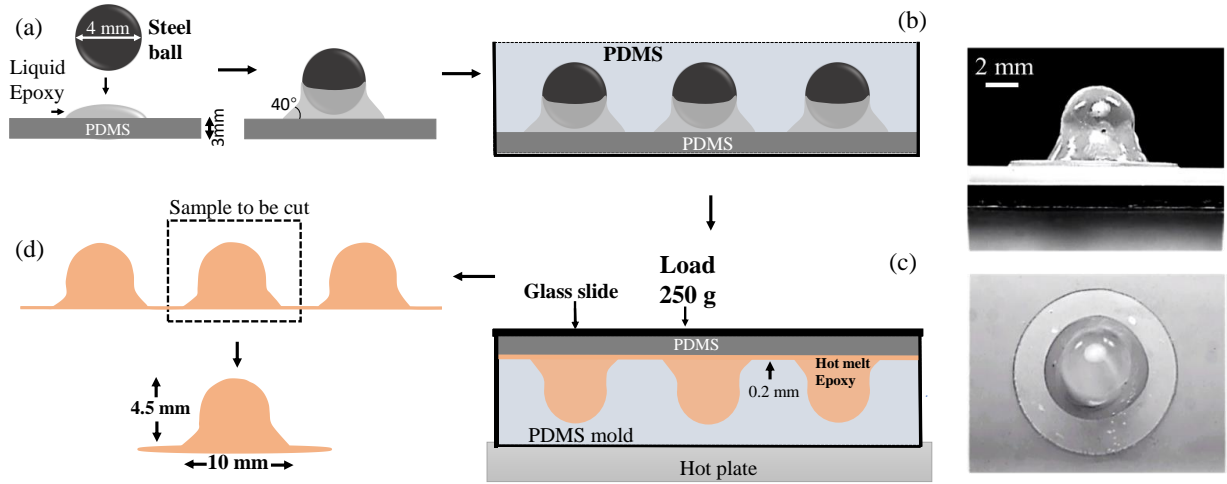

**Figure 5,** Preparation of the domed disk-shaped object. To begin with, a steel ball is brought into contact with a small drop of liquid epoxy, which partially wets the steel ball (a). After the epoxy is cured, the assembly is encased inside cross-linkable PDMS (b). After the PDMS is fully crosslinked, it was removed from the steel-epoxy assembly. This negative image of the steel-epoxy assembly was then filled with hot melt epoxy (c). After cooling, the epoxy structure is removed from which several domed disks were cut out for the adhesion studies (d). Microscopic images of the front and plan views of the pseudo-foulant adhered to the PDMS substrate are shown on the right side of the Figure 6 B.

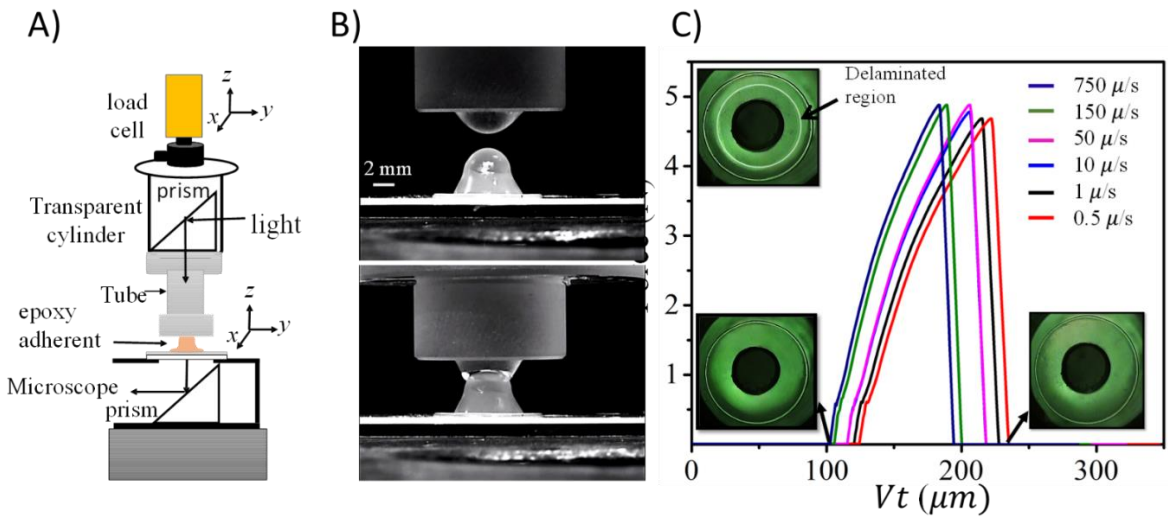

**Figure 6.** The axisymmetric pull-off force apparatus is shown in (A). In order to grab the domed epoxy disc for the pull-off experiment a small amount of the hot melt epoxy deposited on a small screw attached to the load cell was brought into contact with the upper part of the domed disc. As the upper epoxy dome cooled down, it formed a strong bond with the domed epoxy disc below it (B). Pull-out force (C) of the domed disk from the PDMS substrate in air as a function of time,

the inset shows the time stamp images of the growth of delamination from the center to the periphery of the domed disk.

When attempt is made to dislodge the domed disk (elastic modulus = 25 MPa) by uplifting it vertically Figure 6, a substantial force is required with an unmodified PDMS elastomeric film. Data summarized in Figure 6 show that the adhesion force does not depend substantially on the lifting velocity that varied from 0.5 $\mu m/s$ to 750 $\mu m/s$ in these experiments. Furthermore, the total vertical distance (~ 95 $\mu$m) travelled by the adherent before its failure from the substrate is nearly independent of the lifting speed. In all cases, the peak force is achieved when the crack diameter is slightly higher than the central bulky region of contact. While it is somewhat difficult to report a failure stress in these experiments, a nominal stress of about 0.5 MPa is estimated by dividing the peak force with the area of delamination corresponding to the maximum pull-off force. Similar detachment force is observed when the joint is submerged in water for a week, thus illustrating that no deterioration of adhesion takes place in water with an unmodified PDMS elastomer. A different story unfolded with H-PDMS films.

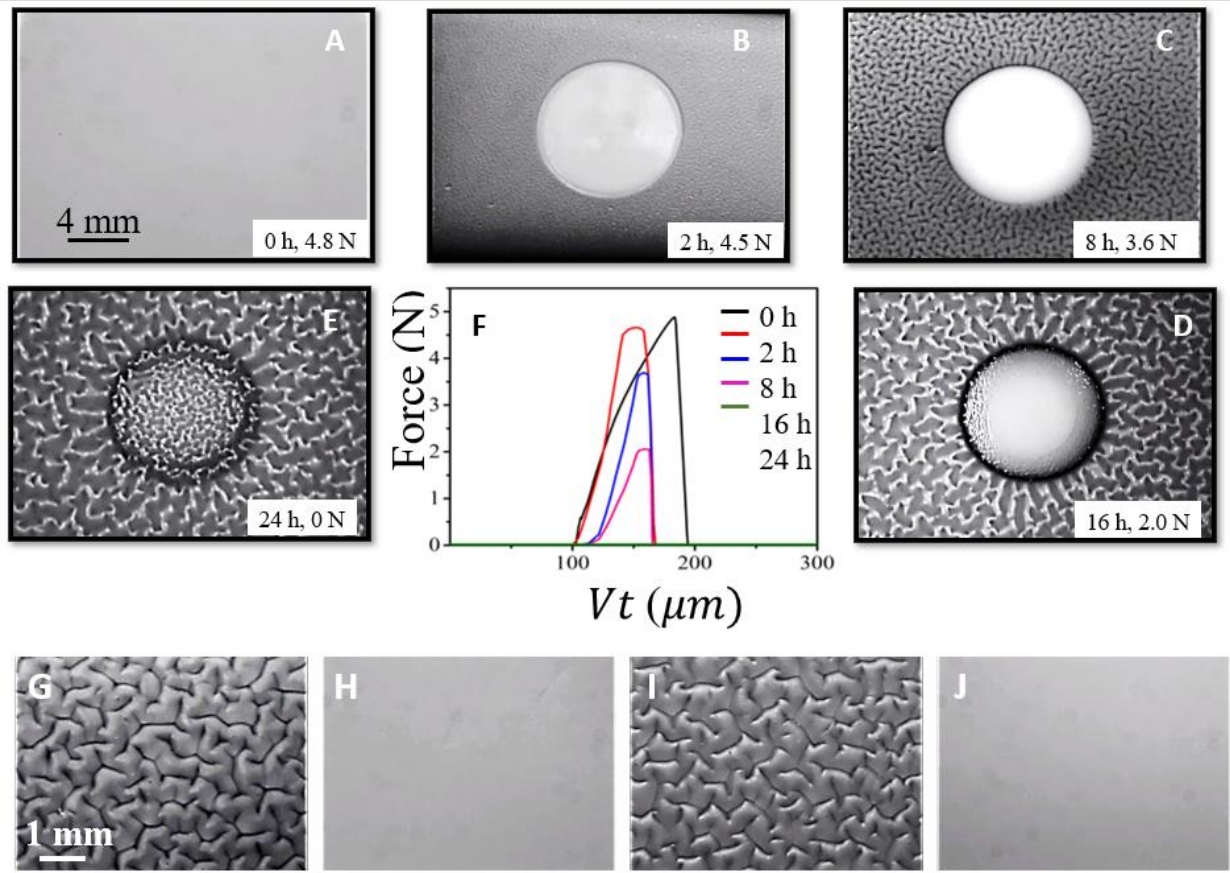

**Figure 7.** (A-E) Micrographs of the delaminated region of a 650 μm thick film of H-PDMS following the removal of a domed disk shaped adherent under water. The pull-off force decreased with the time of immersion. The pull-off forces are stamped inside the micrographs. of the pseudo-foulant. The time dependent force (E) as the domes disk is pulled off the H-PDMS film is plotted as a function of the vertical displacement, which is product of pull-off velocity (150 μm/s) and time. (G-J) These micrographs show that a fully developed instability pattern (G) on a PDMS film (650 μm thick) disappears (H) when it is exposed to dry air (humidity~ 10%). The pattern re-emerges when the film is immersed in distilled water and upon drying it disappears again(J). This sequence exhibits the reversibility of elasto-osmotic stress driven pattern formation on PDMS that can be repeated numerous times.

Here, in Figure 7 the domed disk detaches from the film with a force comparable to that of the unmodified PDMS if the measurement is made soon after the contact is made. Following detachment, the region of original contact remains smooth, i.e. no roughening of the PDMS coating is evident during short contact time. However, if the domed disk is detached after the joint is kept submerged in water for longer time, the detachment force decreases systematically. After 24 h, the domed disk detaches spontaneously from the hygro-elastomeric film without any measurable force. Microscopic observation reveals that fine morphological patterns starts developing on the H-PDMS film underneath the domed disk after it is immersed in water for a short duration. Initially, the patterns are evident near the edge, but it occupies the entire area of contact after prolonged immersion. This suggests that water penetrates the interface from the edge of contact towards the central region.

These experiments, and more to be presented below, insinuate that the detachment of the adherent is mediated by the roughening of the coating. However, there is the possibility that certain components of the hydrophilic inclusion leaches out of the coating and reduces adhesion after accumulating at the interface. In order to address this issue, we immersed H-PDMS coated glass substrates in distilled water and 0.6 M aqueous salt solution, and examined if any leachate is observable in the aqueous phases by analyzing it using UV spectroscopy. A H-PDMS coated glass substrate was also immersed in distilled water and sonicated each day before the aqueous contents were examined. As a control, an equivalent amount of the hydrophilic content present in the coating was mixed directly with the distilled water, with respect to which all the other data were compared. The results summarized in Figure 8 show that the total hydrophilic content leaching out of the coating is much less than that of the control. More leachate was observed when the coated glass slides were sonicated. Furthermore, the contents leaching in salt water seems to be quite negligible. Various studies performed so far indicates that the peak appearing at 210 nm in the salt water (Figure 8 a and b) is due to the enrichment of salt in the solution as the film selectively removes water from it.

**Wettability of the Coating.** We also used another method that provided additional insight into this issue. Since contact angle is sensitive to interfacial processes, we performed an experiment in which a small ( diameter of 3.3 mm, volume of half sphere= 10 mm$^3$ for day 1) air bubble was released over onto a H-PDMS coated glass slide that was pre-immersed in water or salt water at an inclination of 35º from the horizontal plane. The air bubble adhered to the H-PDMS film as soon as it made contact with it. The initial advancing (or recently advanced) and receding (or recently receded) contact angles were 99º and 63º respectively in water; and 99º and 74º in 0.6 M salt solution. The contact angles changed to some degree over three days of immersion in distilled and salt water, reaching to $\theta_a = 103^o$, $\theta_r = 88^o$ in distilled water; and $\theta_a = 105^o$, $\theta_r = 92^o$ in salt water, following which the angles changed more slowly. No detachment of bubbles occurred within seven days in distilled water.

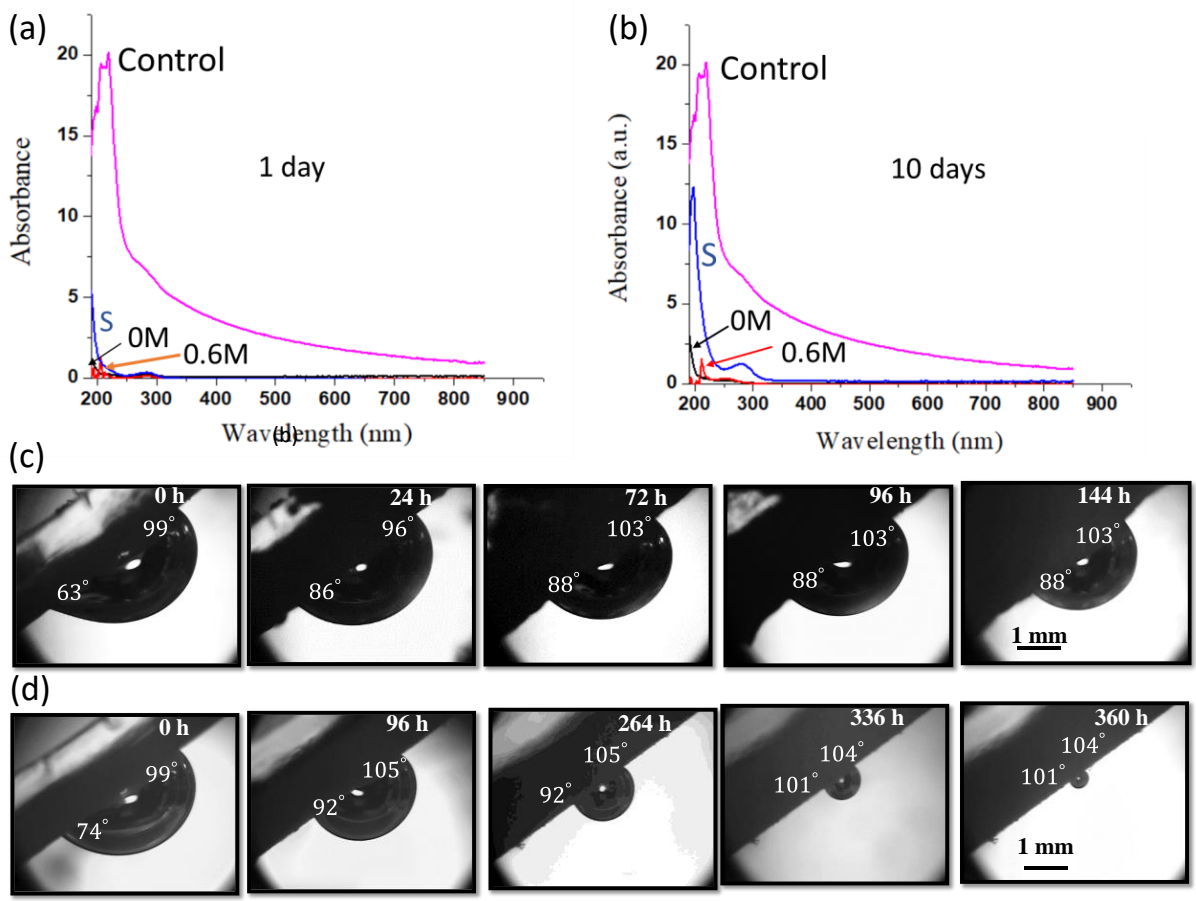

**Figure 8**. UV spectra of the control sample, and the leachates obtained from a sonicated film (S), and unsonicated film (650 μm thick) immersed in water and 0.6 M salt water after one day (a) and seven days (b). Panels (c) and (d) show the fate of air bubbles stuck to a 650 μm thick H-PDMS film as a function of days of immersion in distilled water (c) and salt water (d), respectively.

However, all along, the bubble continued to slide upward, albeit slowly. On the eighth day, the bubble slid all the way upward in distilled water. In salt water, on the other hand, the bubble continued to shrink in size till it disappeared, i.e. the air contained in the bubble was dissolved within the film and/or the surrounding aqueous phase. These experiments coupled with what we learned from UV spectroscopic studies indicate that while small amount of leachate out of the coating may potentially modify the interfacial energetics to some extent, the effect may not strong enough to cause a total loss of interfacial adhesion of the bubble with the coating. A possible scenario may hold for the adhesion of the domed disk with the H-PDMS coating as well. The following experiments support that the increase of roughness due to surface instability may be the primary cause for the reduction of adhesion observed in the experiments described in Figure 7 (A-F).

**Dynamics of Adhesive Delamination Studied with a Foton Probe.** Motivated by the results summarized in Figure 4, we explored the kinetics of the growth of interfacial roughness in the following experiments. Specifically, an optically reflective circular aluminum foil was first attached to the top of the domed disk, which was initially in perfect contact with the H-PDMS coating. After attaching the aluminum foil to the domed disk, the glass slide was kept in a Petri dish following which enough water (or salt water) was added to it.

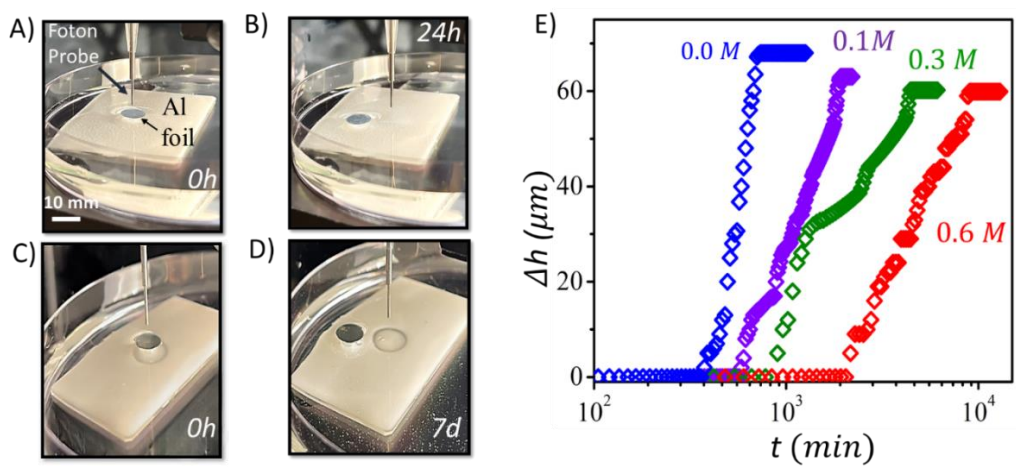

**Figure 9.** (A-D) Microscopic images of domed disk attached to the 650 μm thick H-PDMS film in distilled water (A and B) and 0.6M salt water (C and D). A thin aluminum foil is attached to the upper part of the domed disk, which is immersed in the aqueous phase with the foil touching the liquid-vapor interface. The Foton probe measures the vertical displacement ($\Delta h$) of the foil, thus the domed disk, as the H-PDMS film swells by absorbing water. When these experiments were carried out with a PDMS film without hydrophilic inclusion, $\Delta h$ was undetectable. After the developments of the instability patterns, the domed disk loses complete adhesion to H-PDMS, which is noticeable in that it slides away easily from the original contact region. Development of $\Delta h$ as a function of time depends on the molarity of the salt solution, as shown in Figure (E). Final values of $\Delta h$ are as follows: 68 μm in Distilled water:, : 63 μm in 0.1 M salt solution, : 62 μm in 0.3 M salt solution and : 62 μm in 0.6 M salt solution.

At this stage, the aluminum foil remained at the same level as the air-water interface, while the adhesive interface was totally submerged in the aqueous phase. A Foton probe placed about 2 mm above the coating registered the vertical displacement ($\Delta h$) of the foil as a function of time. As the H-PDMS swelled and the instability developed, the aluminum foil moved vertically outward, which was detected by the Foton probe.

The data summarized in Figure 9 show how the vertical displacement of the film increases in distilled water and salt solutions. As expected, the film swells faster in water than in the salt solutions. After about one day, the domed disk lost complete adhesion with H-PDMS in distilled water. A very slight agitation of water by a pipette caused the adherent to move out of the original contact. As the concentration of salt increases, longer time was needed before the domed disk lost adhesion to the film, with the detachment time increasing with the salinity of water. The maximum vertical displacement of the film reached, more or less, to about 60 μm before the dislodgement of the adherent took place.

The morphological patterns that develop in H-PDMS films are reversible. When pronounced instability pattern develops on a film in pure water after one day of immersion, the film loses the features when exposed to a dry atmosphere Figure 7 H. When such a pattern developed in pure water is immersed in salt water, the amplitude of the pattern decreases from about 62 μm to 42 μm. Then again, when the flattened film is immersed back to distilled water, the morphological patterns re-emerge Figure 7 J fully.

**Consolidation of Main Points.** We reported two main findings. The first one is that a thin film of PDMS with hydrophilic inclusions bonded to a rigid support swells upon adsorbing moisture from the environment, which ultimately creates morphological instability on the surface. Similar instabilities develop when the coatings are exposed to humid atmosphere as well (these results are not reported here). As the driving force for swelling and thus the morphological instability results from the difference of the chemical potential of water in the film and that in the atmosphere, there is an opportunity here to control the nature of the morphological patterns by varying the osmolarity of inclusions and/or the environment.. With the PDMS used here, no short term deterioration of the coating is observed when it is immersed in distilled water for a few days. However, after long immersion, the coating swells so much that it cannot sustain the osmotic stress. In that case, local blisters are formed and some of the hydrophilic content of the coating is released. This is not an issue with a salt solution, where the osmolarity of water in the film and the surrounding phase can ultimately be balanced. However, as the osmolarity cannot be balanced in pure water, the film continue to swell until the hydrostatic stress of the elastic matrix balances it. The matrix has to be deformed to the point that the issue becomes whether the stressed elastomer is strong enough to withstand the osmotic stress. While an elastomer that is more resilient than Syl 184 would be required to solve the issue of long-term durability, we are also considering using crosslinked super-absorbent hydrogel, which would provide an additional control to the developing morphologies owing to its own elastic structure aside from that of the PDMS matrix. At present, we have been able to circumvent the long term deterioration of the coating somewhat by applying a thin film of PDMS (without hydrophilic inclusions) on the top of H-PDMS.

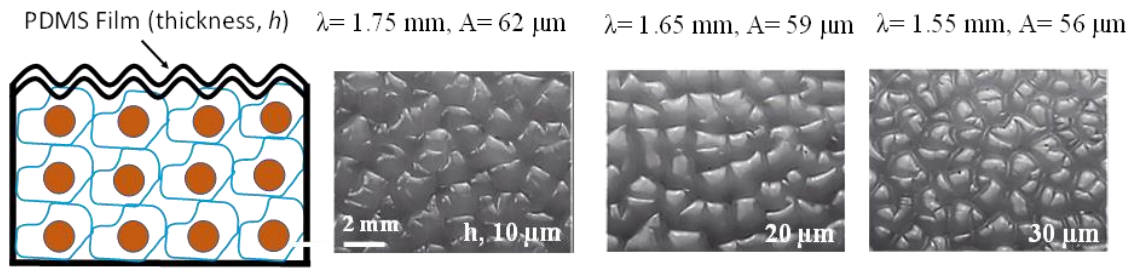

**Figure 10.** Morphological patterns develop even when a H-PDMS is top coated with a thin film of PDMS without any hydrophilic inclusions in it. The patterns developed with three top coats of thickness 10 mm, 20 mm and 30 mm and the corresponding wavelengths and amplitudes are shown here in this figure.

Figure 10 shows that morphological instability still develops in such composite coatings, where the wavelength and the amplitude of the instability is mildly affected by the thickness of the PDMS topcoat the thickness of which varied from 10 $\mu$m to 30 $\mu$m. In a future publication, we wish to report how the stability of the overall structure is improved with such an approach.

The linear dependence of the wavelength of the instability on the film thickness is similar to previous observation where a negative hydrostatic stress develops in the film. However, the proportionality constant between wavelength and thickness is found to be lower than that observed previously. A detailed theoretical analysis is required in order to unravel the reasons for this difference although from the outset, we note that the microscopic elastic property of the H-PDMS film is heterogeneous which differs from the previous studies. Furthermore, as the matrix swells, its elastic property is expected to change rather non-linearly in these highly deformable systems, which would also have to be considered in the detailed analysis of the problem.

We also provided a proof of the concept that the morphologies that develop in H-PDMS pushes out an adherent attached to it both in water and salt water to the extent that total loss of adhesion is feasible. Some of the leachates may aid the release process; but it may not be entirely needed to meet our objective. One of the main goals of future studies is to increase the permeability of H-PDMS in order to decrease the response time for auto-failure of an adherent sticking to it. This could be achieved with a block copolymer consisting of hydrophilic and hydrophobic segments.[39]

**Conclusion** A thin film of hygroscopic elastomer bonded to a rigid substrate undergoes a morphological instability, which can be used to dislodge an adherent attached to it. While more studies are needed to optimize the properties of such a film, we feel this approach to control release as expounded here has potential.

**Experimental Methods**

**Materials.** The materials for the synthesis of *adhesion primer* are as follows: bis (triethoxysilyl) ethane, vinyl trimethoxy-silane, Titanium n-Butoxide (TBT), and Platinum-Divinyltetramethyldisiloxane complex in xylene (Karstedt's catalyst) (all these chemicals were purchased from Gelest Inc., USA); hydrogen silsesquioxane (HSQ, H-resin, Dow Corning, USA); Chloroform (Carolina Biological Supply, USA). Clean rectangular glass slides (7.5 cm x 5 cm x 1 mm) were purchased from Fisher Scientific, USA. The materials for the *synthesis of H-PDMS* are as follows: Polydimethyl siloxane (PDMS), which was prepared using Sylgard$^{TM}$ 184 (silicone elastomer base & Curing agent, purchased from Solar Panel Encapsulator, Trademark of The Dow Chemical Company, Virginia, sold by Amazon USA ); silicone glycol (Q2-5211, Dow Corning, USA); Poly(acrylic acid) or PAA (average MW= 3,000,000, Sigma Aldrich, USA); Polyvinyl Alcohol, or PVA ( average Mw= 125000, Aldrich Chemical Company); glycerol (99+% pure, Sigma-Aldrich, USA); D-Glucose (anhydrous, Fisher Scientific, USA); and D-Fructose (99%) (Alfa Aesar, USA). Additional chemicals were: Salt (Sea Salt purchased from Instant Ocean); hot melt epoxy (Surebonder, USA) and clear weld epoxy ( J-B Weld, USA). Water was distilled an deionized (resistance =18 M Ω).

**Analytical and Process instruments** The pull-off force of a domed disk shape adhering to H-PDMS film was measured using a load cell (model L2338, Futek Advanced Sensor Technology, Inc., Irvine, CA, USA) assisted by an Arduino module. A motorized vertical stage (Melles Griot Photonics Components, Carlsbad, CA, USA) allowed the load cell to move at various speeds. The mechanism of release was visualized using an optical microscope (Nikon, model SMZ-2T, Mager Scientific, Dexter, MI, USA) equipped with a CCD video camera (Sony, model XC-75, Optical Apparatus Co., Ardmore, PA, USA) and a video recording system. The vertical deflection of the H-PDMS film immersed in water was measured as a function of time using a Foton probe (MTI-

2100 Fotonic sensor purchased from MTI Instruments, Inc., USA) this was interfaced with computer using a NIDAQ Data Acquisition system (analog to digital data acquisition unit (Model – USB 6003) purchased from National instruments). The turbidity measurements were performed occasionally to test the quality of water using a turbidity meter (Orion AQUAfast AQ3010), Thermo Fisher Scientific, USA). Further analysis of water in which H-PDMS was submerged was carried out using UV-Visible spectrophotometer (ND-ONEC-W4) purchased from Thermo Fisher Scientific, USA.  The spin coater used to deposit a primer on the glass slide was purchased from Headway Research Inc. USA.

H-PDMS films of controlled thickness were prepared on glass slides using a standard Film Casting Doctor Blade. A heat gun (Craftzman, USA) used for different heating purposes (e.g. see below) was purchased from a local vendor (home depot). A sonicator ( model FS 5) used for certain experiment was purchased from Fisher Scientific, USA. All the glass containers were purchased from Fisher scientific, USA.

**Preparation of the pseudo foulants.** Preparation of the dome shaped disk adherent required several steps as follows. First, a PDMS substrate used to prepare the adherent was made by mixing a 10:1 ratio of sylgard 184 elastomeric base and the cross linker. After degassing the above mixture under gentle vacuum for half an hour, it was transferred to a polystyrene petri dish, cured in an oven at 70° for 1 h.  A steel ball (diameter 4 mm) was gently placed on a (5 µl) drop of a clear weld epoxy (1:1 mixture of resin and hardener) that was pre-deposited on a PDMS substrate. The PDMS substrate containing several such steel balls was left undisturbed overnight to ensure that the epoxy adhesive was fully solidified, following which, it was  encapsulated with degassed Sylgard-184 and heated to 60 °C for 3 hrs. The negative image of the steel ball-epoxy structure as transferred to PDMS comprised of dimples where the steel balls were.  Each  dimples was then filled with 0.12 g of a hot melt epoxy and covered with a rigid glass plate, following which it was pressed down with an additional 250 g weight. The mold was heated gradually on the hot plate for 10 minutes until the temperature reached 250 $^0$C . This procedure re-melted the epoxy thus allowing it flow and conform to the dimpled PDMS. Where the liquid adhesive merged (outside the dimples), it formed a thin continuous film. After allowing the structure to cool down to room temperature, the epoxy mold was gently separated from PDMS. Finally, a sharp stainless-steel needle was used to cut out a circular domed disc using a motorized shaft rotating at a speed of 50 RPM.

**Preparation of the primer.** The primer was prepared by mixing the following components (given in terms of weight percent in chloroform): HSQ (3.3 %), bis (triethoxysilyl) ethane (0.2 %), vinyl trimethoxy silane (0.9 %), TBT (0.2 %), and platinum catalyst (0.033 %), which produced a clear colorless liquid. Adding large amounts of solutes in the solvent can result in gelation, which was avoided for the current studies.

**Preparation of sugar solution.** The concentrated sugar solution was prepared by mixing aqueous solutions of D-Fructose (31.5% w/w) and D-Glucose 10.5% (w/w). The mixture was heated to 65 °C with constant stirring until a homogeneous solution was obtained. To this solution 10 ppm (by weight) of sodium azide was added that prevents growth of bacteria. The final solution was stored in a refrigerator, but was brought to room temperature before further use.

**Preparation of H-PDMS Films.** A mixture of the sugar solution and polyacrylic acid was prepared by mixing 3.7 % (w/w) of PAA in the sugar solution until PAA was dissolved in it homogeneously. The composite coating was made by mixing 36% (w/w) of the above sol-gel solution with, 7 % (w/w) of glycerol, 1 % (w/w) of superwetting agent, and 56 % (w/w) of PDMS

& its corresponding quantity of its cross linker (10:1.2). While most of the experiments were performed with polyacrylic acid, we have also used polyvinyl alcohol occasionally. With the latter, 7% of an aqueous solution of PVA was added to the sugar solution and was heated with continuously stirring at 60C° till a clear solution was formed. 36% of this solution was then blended with the other components to prepare the base of the composite coating.

First, The sol-gel and glycerol were mixed very well for an hour until a homogeneous clear mixture was prepared, following which PDMS elastomer and the superwetting surfactant were added to it. A four-blade agitator connected to a rotating (800 RPM) shaft electric motor was used for blending the above mixture for two hours, which resulted in a white viscous material with a viscosity of 58 poise.

**Bonding of H-PDMS to Glass Slide.** The primer was coated on flame treated glass substrates using a spin coater with a speed of 1000 rev/min, which resulted in a 3 µm thick primer. We found that this primer can be used soon after drying, e.g., after ten minutes or even after one or two days of leaving them in atmospheric condition. The composite coating was then coated on the top of the primer layer after it dried using the Doctor Blade instrument that resulted in leveled coatings with the desired thickness. The samples were stored in atmospheric conditions for 5 hours before they were cured in the oven at 70 °C for 2 hours. Various thicknesses of the coating were prepared following the above method.

**Investigation of Morphological Instability.** The H-PDMS films of six different thicknesses, 50 µm to 700 µm, were soaked in distilled water and three different concentrations of aqueous salt solution, i.e., 0.1 M, 0.3 M and 0.6 M and pure distilled water. The surface instability patterns were from the images taken with a INFINITY photo-optical inverted microscope that was connected to a CCD camera and a video recorder. Five images were taken by translating the microscope to different positions using horizontal and vertical stages. The results of image characterization of the same sample were averaged for more accurate results.

**Analysis of Leachate using UV Spectroscopy.** In order to investigate if any of the components of the hydrophilic inclusion is leached out of the H-PDMS films, three experiments were conducted for a 650μm thickness H-PDMS coating of a mass of 2.6g on a glass substrate (7.5 x 5 cm) after cleaning it properly using distilled water. In the first experiment, the sample was immersed in 30 ml distilled water inside a petri-dish, and it was covered and sealed with parafilm to avoid any evaporation. The same procedure was followed for experiment 2 as experiment 1 except that 30ml of 0.6M salt water was used instead of distilled water. In the third experiment, the H-PDMS coating was removed from the glass substrate and cut into small squares~(5 x 5mm); the cut coating was immersed in 30ml distilled water inside a glass vail and it was sonicated using a sonicator to find the maximum inclusion leached out. The water used for immersing the samples in the above three experiments was mixed before taking 3 ml sample for investigation analysis each 24 h for 10 days using UV-Visible spectrophotometric. The sonicated sample was sonicated for 3 hours daily before doing the analysis. For qualitative and quantitative analysis, the results were compared with a reference, control experiment, in which similar amount of the hydrophilic inclusion in 2.6g of H-PDMS was mixed with 30 ml water and analyzed using UV-Visible spectrophotometer.

Another qualitative experiment was conducted using Benedict's solution qualitative for detection of reducing sugar by color change in case of the presence of the reduced sugar.

**Pull-off Force Measurement.** After placing the domed epoxy disks on the H=PDMS films, they were subjected to a gentle vacuum (30 Hg) for half an hour to remove bubbles from the adhesive junction. Following this treatment, the domed disk was re-melted with a brief (2 sec) exposure to the hot air of a heat gun at 130 C°. After the samples were cooled, pull-off forces were measured. A detailed description of the pull-off force apparatus is described in reference.[40]

**The Axisymmetric Pull-off force Apparatus** The axisymmetric pull-off force apparatus used in these studies (Figure 6 A) is similar to that described in the past. The main issue here was to grab the domed epoxy disc appropriately so that the pull-off experiment could be carried out. For this purpose, a screw was attached to the bottom of the load cell, which had a dimple at the end, A small amount of the hot melt epoxy was first deposited on it, which formed the shape of a dome. As the load cell could be moved in horizontal and vertical directions using translation stages, it was positioned so that the epoxy on the tip of the screw comes close to the epoxy adherent below it. The upper epoxy dome was melted with a heat gun, which was quickly brought into contact with the domed disc below it. As the upper epoxy dome cooled down, it formed a strong bond with the domed epoxy disc below it. The load cell is connected to a motorized vertical that allowed measurement of the pull-off force at different speeds. The force was measured by the load cell (sensor) digitally using an Arduino module. The mechanism of release was observed using optical microscopes equipped with a CCD video camera. The self-explanatory schematic of the optical path to observe the adhesive interface is shown in Figure 6 A.

Investigation the vertical deflection of the instability

In this experiment, the vertical deflection distance (amplitude) was measured as a function of time using a Fotonic sensor. A flexible fiber-optic probe is used by the Fotonic Sensor to emit a beam of light, which is then reflected back from the target surface. The sensor measures the distance between the probe tip and the target by calculating the time it takes for the light to travel from the sensor to the target and back. The resulting electrical signal is proportional to the distance between the probe tip and the target being measured. The experiment was done as follows; An optically reflective circular aluminum foil (target surface) was affixed to the top of the domed disk, which was initially in perfect contact with the H-PDMS coating. The target surface on a glass slide was then placed in a petri dish, and a photonic probe was positioned 2 mm above it. The H-PDMS coating was submerged in water (or salt water), causing it to swell and the instability developed and grew with time, As a result of this instability, the aluminum foil was displaced vertically outward, and this displacement was detected by the Foton sensor. The data acquisition continued until the domed disk completely dislodged from the coating. The time for dislodge of the disk was monitored by creating gentle agitation in the bulk of the liquid using a sterile pipette to force air to create a flow of liquid in the petri-dish. When the domed-disk was no more in contact with the coating, it started to float away with the gentle flow created above. At this stage, the data acquisition of the Foton probe was truncated and the coating was analyzed using microscopy to cross-verify the thickness of the instability measured using this technique and image analysis using Gwyddion (described below).

**Fast Fourier Transformation (FFT)** The micrographs were first processed with ImageJ software, where the images was calibrated to its scale, and the background noise was removed by RGB (what is RGD) filter. Further analysis on these images were performed using a Gwyddion V 2.60 software. The filtering procedure was based on Gwyddion software[38] , which is an Open Source project. It is one of the widely used freeware programs, as it provides a wide spectrum for analysis and data representation tools. This software is purposely designed to analyze the height fields of structures captured via micrographic methods. The 3D amplitude of the instability on these coatings was also estimated using this method by setting the cross-section of the micrograph of the coating as a reference. After fixing the top horizontal position of the pre-soaked coating as origin (0,0,0), the amplitude scan was performed using the extract profiles program[38] (using the lines tool) along the horizontal on the surface of the hydrated coatings. A line profile analysis was undertaken using the profile extract function across the section on all the images to quantify the periodic and morphological properties of the thin film coatings. The open architecture of the software also allowed to implement the filtering feature based on the wavelet transform and inverted wavelet transform of the two-dimensional set of data acquired using microscopy. Hence, the fast Fourier transform (FFT) of the instabilities developed on the coatings were estimated using the above tool. These signals were decomposed into mutually orthogonal set of wavelets, which was averaged over the entire scale of the micrograph to obtain the wavelength of the instabilities developed on the coatings.

Further, in-order to compute the power spectrum of the coatings, the FFT was also analyzed using a m-script code in MATLAB R2017b. Here, the 'rgb2gray' function was used to convert the 2D micrographs into grayscale. The script uses the 'imread' function to upload the image into the MATLAB environment. The 'imshow' function was used to plot the original image. To transform the images into the frequency domain, 2D fast Fourier transform (FFT) algorithm was used, in which, the power spectrum of the micrograph in the imaginary space were obtained using the following equation: $z(x) = |f(x)|^2$. Where, $f(x)$ is the Fourier transform of $f(v)$,, $f(v)$ being is wave vector of the 2D micrograph in the frequency domain. FFT of a micrograph, with filtered background noise, consists of waveforms of different frequencies which is integrated to get the original micrographic image. fft2 function in MATLAB was used to perform the 2D FFT on these images [4]. The advantage with this method of pre-processing the image by removing the background noise and enhancing the contrast in imageJ helped to obtain a 8-bit scale of the final images. The final result of this FFT was viewed with $q = fftshift(x)$. The data obtained by using gwyddion and matlab showed same analytical results.

**References**

1. Banerjee, I., Pangule, R. C., & Kane, R. S. Antifouling coatings: recent developments in the design of surfaces that prevent fouling by proteins, bacteria, and marine organisms. *Advanced materials*, **2011,** *23*(6), 690-718.
2. Fitridge, I., Dempster, T., Guenther, J., & De Nys, R. The impact and control of biofouling in marine aquaculture: a review. *Biofouling*, **2012,** *28*(7), 649-669.
3. Brady Jr, R. F., & Singer, I. L. Mechanical factors favoring release from fouling release coatings. *Biofouling*, **2000,** *15*(1-3), 73-81.
4. Wendt, D. E., Kowalke, G. L., Kim, J., & Singer, I. L. Factors that influence elastomeric coating performance: the effect of coating thickness on basal plate morphology, growth, and critical removal stress of the barnacle Balanus amphitrite. *Biofouling*, **2006,** *22*(1), 1-9.
5. Chaudhury, M. K., Finlay, J. A., Chung, J. Y., Callow, M. E., & Callow, J. A. (2005). The influence of elastic modulus and thickness on the release of the soft-fouling green alga Ulva linza (syn. Enteromorpha linza) from poly (dimethylsiloxane)(PDMS) model networks. *Biofouling*, **2005,** *21*(1), 41-48.
6. Shull, K. R. Contact mechanics and the adhesion of soft solids. *Materials Science and Engineering: R: Reports*, **2002,** *36*(1), 1-45.

7. Creton, C., & Ciccotti, M. Fracture and adhesion of soft materials: a review. *Reports on Progress in Physics*, **2016,** *79*(4), 046601.

8. Mukherjee, R., & Sharma, A. Instability, self-organization and pattern formation in thin soft films. *Soft matter*, **2015,** *11*(45), 8717-8740.

9. Chaudhury, M. K., Chakrabarti, A., & Ghatak, A. Adhesion-induced instabilities and pattern formation in thin films of elastomers and gels. *The European Physical Journal E*, **2015,** *38*, 1-26.

10. Schmidt, D. L., Brady, R. F., Lam, K., Schmidt, D. C., & Chaudhury, M. K. Contact angle hysteresis, adhesion, and marine biofouling. *Langmuir*, **2004,** *20*(7), 2830-2836.

11. Shivapooja, P., Wang, Q., Orihuela, B., Rittschof, D., López, G. P., & Zhao, X. Bioinspired surfaces with dynamic topography for active control of biofouling. *Advanced Materials*, **2013,** *25*(10), 1430-1434.

12. Liu, J., Yang, Z., Wen, G., Wang, Z. P., & Xie, Y. M. Contact mechanics model of wrinkling instability of dielectric elastomer membranes for anti-biofouling. *Materials Today Communications*, **2023,** *34*, 105216.

13. Bogy, D. B., & Wang, K. C. Stress singularities at interface corners in bonded dissimilar isotropic elastic materials. *International Journal of Solids and Structures*, **1971,** *7*(8), 993-1005.

14. Gorb, S., Varenberg, M., Peressadko, A., & Tuma, J. Biomimetic mushroom-shaped fibrillar adhesive microstructure. *Journal of The Royal Society Interface*, **2007,** *4*(13), 271-275.

15. Mönch, W., & Herminghaus, S. Elastic instability of rubber films between solid bodies. *Europhysics Letters*, **2001,** *53*(4), 525.

16. Vorvolakos, K., & Chaudhury, M. K. The Role of Interfacial Slippage in Adhesive Release. *ACS Symposium Series,* **2000,** Vol. 741, pp 83-90.

17. Chaudhury, M. K., Vorvolakos, K., & Malotky, D. (2008). Friction at soft polymer surface. In Polymer Thin Films, Series in Soft Condensed Matter; World Scientific, **2008,** Vol. 1, pp 195–219.

18. Mazurek, P., Hvilsted, S., & Skov, A. L. Green silicone elastomer obtained from a counterintuitively stable mixture of glycerol and PDMS. *Polymer*, **2016,** *87*, 1-7.

19. Borde, A., Larsson, M., Odelberg, Y., Hagman, J., Löwenhielm, P., & Larsson, A. Increased water transport in PDMS silicone films by addition of excipients. *Acta Biomaterialia*, **2012,** *8*(2), 579-588.

20. Style, R. W., Boltyanskiy, R., Allen, B., Jensen, K. E., Foote, H. P., Wettlaufer, J. S., & Dufresne, E. R. Stiffening solids with liquid inclusions. *Nature Physics*, **2015,** *11*(1), 82-87.

21. Moser, S., Feng, Y., Yasa, O., Heyden, S., Kessler, M., Amstad, E., & Style, R. W. (2022). Hydroelastomers: soft, tough, highly swelling composites. *Soft Matter*, **2022,** *18*(37), 7229-7235.

22. Kuhar, K., Jesbeer, M., & Ghatak, A. Soft Gel-Filled Composite Adhesive for Dry and Wet Adhesion. *ACS Applied Polymer Materials*, **2021,** *3*(8), 3755-3765.

23. Kuhar, K., & Ghatak, A. Two-Phase Composite Adhesive with Viscoelastic Inclusion. *ACS Applied Polymer Materials*, **2022,** *4*(12), 9095-9102.

24. Park, Y. L., Majidi, C., Kramer, R., Bérard, P., & Wood, R. J. Hyperelastic pressure sensing with a liquid-embedded elastomer. *Journal of micromechanics and microengineering*, **2010,** *20*(12), 125029.

25. Markvicka, E. J., Bartlett, M. D., Huang, X., & Majidi, C. An autonomously electrically self-healing liquid metal–elastomer composite for robust soft-matter robotics and electronics. *Nature materials*, **2018,** *17*(7), 618-624.

26. Jha, A., Karnal, P., & Frechette, J. Adhesion of fluid infused silicone elastomer to glass. *Soft Matter*, **2022,** *18*(39), 7579-7592.

27. Wang, J., Wu, B., Dhyani, A., Repetto, T., Gayle, A. J., Cho, T. H., & Tuteja, A. (2022). Durable Liquid-and Solid-Repellent Elastomeric Coatings Infused with Partially Crosslinked Lubricants. *ACS Applied Materials & Interfaces*, **2022,** *14*(19), 22466-22475.

28. Kabiri, K., Omidian, H., Hashemi, S. A., & Zohuriaan-Mehr, M. J. Synthesis of fast-swelling superabsorbent hydrogels: effect of crosslinker type and concentration on porosity and absorption rate. *European Polymer Journal*, **2003,** *39*(7), 1341-1348.

29. Baker, H. G., Baker, I., & Hodges, S. A. Sugar composition of nectars and fruits consumed by birds and bats in the tropics and subtropics 1. *Biotropica*, **1998,** *30*(4), 559-586.

30. Biswas, S., Chakrabarti, A., Chateauminois, A., Wandersman, E., Prevost, A. M., & Chaudhury, M. K. Soft lithography using nectar droplets. *Langmuir*, **2015,** *31*(48), 13155-13164.

31. Song, W., Xin, J., & Zhang, J. One-pot synthesis of soy protein (SP)-poly (acrylic acid) (PAA) superabsorbent hydrogels via facile preparation of SP macromonomer. *Industrial crops and products*, **2017,** *100*, 117-125.

32. Velankar, S. S., Lai, V., & Vaia, R. A. Swelling-induced delamination causes folding of surface-tethered polymer gels. *ACS applied materials & interfaces*, **2012,** *4*(1), 24-29.

33. Chaudhury, M. K., & Gaul, J. H. (**1988**). *U.S. Patent No. 4,737,562*. Washington, DC: U.S. Patent and Trademark Office.

34. Plueddemann, E. P., & Plueddemann, E. P. Nature of adhesion through silane coupling agents. *Springer US,* **1991,** (pp. 115-152).

35. Wang, Q., & Zhao, X. A three-dimensional phase diagram of growth-induced surface instabilities. *Scientific reports*, **2015,** *5*(1), 8887.

36. Klapetek, P., Necas, D. and Anderson, C. Gwyddion user guide. *Czech Metrology Institute*, **2004**, 2007 : p.2009.

37. Poon, T. C., & Banerjee, P. P. Contemporary optical image processing with MATLAB. *Elsevier,* **2001.**

38. Nečas, D., & Klapetek, P. Gwyddion: an open-source software for SPM data analysis. *Open Physics*, **2012,** *10*(1), 181-188.

39. Vaidya, A., & Chaudhury, M. K. Synthesis and surface properties of environmentally responsive segmented polyurethanes. *Journal of Colloid and Interface Science*, **2002,** *249*(1), 235-245.

40. Chung, J. Y., & Chaudhury, M. K. Soft and hard adhesion. *The Journal of Adhesion*, **2005,** *81*(10-11), 1119-1145.